\documentclass[12pt,preprint]{aastex}
\usepackage{lscape}

\begin{document}

\def\esc{{\rm esc}}
\def\cut{{\rm cut}}
\def\tot{{\rm tot}}
\def\thresh{{\rm thresh}}

\newcommand{\kms}{\, {\rm km\, s}^{-1}}
\newcommand{\cm}{\, {\rm cm}}
\newcommand{\gm}{\, {\rm g}}
\newcommand{\erg}{\, {\rm erg}}
\newcommand{\kel}{\, {\rm K}}
\newcommand{\kpc}{\, {\rm kpc}}
\newcommand{\mpc}{\, {\rm Mpc}}
\newcommand{\seg}{\, {\rm s}}
\newcommand{\kev}{\, {\rm keV}}
\newcommand{\hz}{\, {\rm Hz}}
\newcommand{\etal}{et al.\ }
\newcommand{\yr}{\, {\rm yr}}
\newcommand{\mpyr}{{\rm mas}\, {\rm yr}^{-1}}
\newcommand{\msun}{\rm M_{\odot}}
\newcommand{\gyr}{\, {\rm Gyr}}
\newcommand{\eq}{eq.\ }

\title{Old-Population Hypervelocity Stars from the Galactic Center: Limits from the SDSS}
\author{Juna A. Kollmeier\altaffilmark{1}, Andrew Gould\altaffilmark{2}, Gillian Knapp\altaffilmark{3}, and Timothy C. Beers\altaffilmark{4}}

\altaffiltext{1}{Observatories of the Carnegie Institution of Washington,
  813 Santa Barbara Street, Pasadena, CA 91101}
\altaffiltext{2}{Department of Astronomy, The Ohio State University, 4051 McPherson Laboratory, Columbus, OH, 43210}
\altaffiltext{3}{Department of Astrophysical Sciences, Princeton University,
  Peyton Hall, Princeton, NJ 08544}
\altaffiltext{4}{Department of Physics \& Astronomy; CSCE:Center for the Sudy of Cosmic Evolution and JINA: Joint Institute for Nuclear Astrophysics, Michigan State University, E. Lansing, MI 48824 USA}
\begin{abstract}
  We present limits on the ejection of old-population HVS from a
  sample of over 290,000 stars selected from the SDSS.  We derive the
  speed at the solar circle from the measured positions and radial
  velocities by assuming a radial orbit and adopting a simple
  isothermal model of the Galactic halo, which enables us to identify candidate bound and
  unbound ejectees.  We find 4 candidate bound F-stars from this
  sample, all with negative Galactocentric radial velocity
  (i.e., returning toward the GC).  We additionally find 2 candidate
  unbound stars (one F and one G), however, existing proper motion
  measurements make these unlikely to be emerging from the GC.  These
  data place an upper limit on the rate of ejection of old-population
  stars from the GC of $\sim 45\, {\rm Myr}^{-1}$.  Comparing to the rate for
  more massive B-star ejectees of $\sim 0.5\, {\rm Myr}^{-1}$, our limit
  on the rate of ejection of old-population HVS shows that the mass
  function at the GC is not bottom-heavy and is consistent with being
  normal.  Future targeted surveys of old-population HVS could
  determine if it is indeed top-heavy.
\end{abstract}
\section{Introduction}
The search for hypervelocity stars (HVS) in the stellar halo was
originally proposed in order to indirectly probe the Galactic Center
(GC) at optical wavelengths \citep{hills88}.  These stars, having been
ejected from the GC via binary-exchange collisions with a putative
supermassive black hole at speeds in excess of Galactic escape would
provide the dynamical ``smoking gun'' evidence for such an object at a
time when direct detection seemed remote.  Despite the tremendous
advances in infrared instrumentation in the intervening two decades
(and the associated confirmation of the supermassive black hole at the
GC), it remains extraordinarily difficult to see any but the
brightest, youngest stars at the center of the Galaxy
\citep[e.g.][]{genzel97,ghez98, ghez00, schoedel02}.  The existence of
these young stars is a major mystery since one would naively expect
that the conditions near a supermassive black hole are sufficiently
hostile to prevent in situ star formation of any kind.  A key question
is whether or not these stars represent the ``tip of the iceberg'' in
an otherwise normal star forming region or whether star formation is
fundamentally different at the GC.  Because HVS are relatively rare,
it was previously not feasible to amass sufficiently large numbers of
such stars to go beyond the ``smoking-gun'' prediction that there
should be young, blue stars in the halo.  In the era of deep,
wide-field spectroscopic surveys, it is now conceivable to use Hills'
original idea to exploit HVS in the Galactic halo to obtain,
among other things, detailed information about star formation at the
GC, which is otherwise obscured from direct view, to at least a factor
of 2.5 lower in mass than the young population currently being probed \citep{brown05,brown08}.

The Sloan Digitial Sky Survey (SDSS) provides an exceptional database
to search for HVS that would otherwise be missed from targeted
surveys.  The SDSS has obtained over 300,000 stellar spectra since it
began operations \citep{york00, gunn98, gunn06}.  The data are reduced
by an automated pipeline that produces reliable spectral
classifications and radial velocities (RVs).  The SEGUE Stellar
Parameter Pipeline (hereafter, SSPP) further processes the wavelength-
and flux-calibrated spectra generated by the standard SDSS
spectroscopic reduction pipeline \citep{stoughton02}, obtains
equivalent widths and/or line indices for 82 atomic or molecular
absorption lines, and estimates $T_{\rm eff}$, log $g$, and [Fe/H]
through the application of a number of approaches \citep{lee08a,
lee08b, allendeprieto08}.  With this enormous database of stellar
spectra it is already possible to detect HVS among a broad class of
stars. In particular, the size of the SDSS stellar library makes SDSS
sensitive to F and G type HVS which require a substantially different
search strategy than their O and B counterparts
\citep{kandg07}.  In contrast to the photometric survey, the SDSS RV
catalog is not complete in any dimension.  In this work, we examine a
large sample of high-velocity stars in order to place limits on the
ejection of F/G stars from the GC.

In \S~\ref{sec:sdss} we will summarize the stellar spectra and the
sample selection for stars that will be used for this analysis.  In
\S~\ref{sec:ejectees} we present limits on the ejecton of
old-population HVS. We compare this limit with young-population HVS in
\S~\ref{sec:comp_b_stars}.  We present our discussion and conclusions
in \S~\ref{sec:disc}.

\section{{The Sample} 
\label{sec:sdss}}

The sample used here is part of an ongoing study to find high-velocity
stars within the SDSS (Knapp et al. 2009, in preparation).  All RVs
for stars aquired by the SDSS as of 15 January 2007 were extracted
from the SDSS database.  Selected objects were required to have RV errors
not exceeding 100 $\kms$ and required to be free of pipeline flags
indicating likely problems in the automated velocity determination.
This yielded a sample of 291,111 objects.  Velocities were converted
to Galactocentric velocity assuming that the local rotation speed is
220$\kms$ and that the Sun moves at (+9,+12,+7)$\kms$ relative to the
local standard of rest in the direction of the GC, Galactic rotation
and the north Galactic pole respectively.  All objects with
Galactocentric RVs in excess of $\vert V_G \vert \geq 350 \kms$ were
examined to ensure the sample was free of catastrophic velocity
errors.  This process yielded a total sample of 33 objects with
Galactocentric RVs exceeding $\pm 400 \kms$, which we analyze in detail
below.

\section{{F and G Star Ejectees} 
\label{sec:ejectees}}

\citet{kandg07} argued that large RV samples, such as SDSS, could
probe an as yet undiscovered old population of stars ejected from the
GC and would be most sensitive to stars near the
main-sequence turnoff within this population.  Moreover, they argued
that such surveys would be more sensitive to the bound rather than
unbound members of this population, simply because the bound stars
accumulate over the lifetime of the Galaxy and can be seen still
orbiting, while the unbound stars can only be viewed during their
exit.  In this section, we analyze the sensitivity of SDSS to both
classes of ejectees, and we tabulate the candidates derived from this
database.

\subsection{{Selection Criteria for Bound Ejectees} \label{sec:boundcand}}

We begin by asking what the sensitivity of the SDSS
survey is to bound turnoff stars and whether any such candidates
are in the sample.  Of course, this requires that we establish
selection criteria that remove the vast majority of contaminants
but still retain significant sensitivity to bound ejectees.
We adopt the following two criteria:
\begin{equation}
v_{r,G}> v_\cut= 400\,\kms,\qquad v_{\odot\rm -circle} < v_\esc,
\label{eqn:sel_crit}
\end{equation}
where $v_{r,G}$ is the observed RV converted to the
Galactic frame, $v_{\odot\rm -circle}$ is the velocity that
the star has when it crosses the solar circle, and $v_\esc$ is
the Galactic escape velocity, again measured at the solar circle.

The first criterion is purely observational (apart from the implied
assumption of solar motion) and is self-contained.  The threshold
$v_\cut$ is established empirically from the observed velocity
distribution (Knapp et al. 2009, in prep).  The second criterion
requires that we specify both a model of the Galaxy (to determine the
value of $v_\esc$) and a model of the photometric properties of the
target population (ejectees), so that we can estimate $v_{\odot\rm
  -circle}$ from its observed $v_{r,G}$ and from the observed fluxes
in several bands.

For the Galactic model, we adopt an isothermal sphere characterized
by a rotation speed $v_c=220\,\kms$, truncated at a Galactocentric
radius $R_b$.  For this model $R_b = R_0\exp[(v_\esc/v_c)^2/2-1]$.
We will adopt $v_\esc = 550\,\kms$, which implies $R_b = 8.37\,R_0$,
as a default, but will also consider other values.
We will consider two classes of stars,' ``F stars'', defined observationally
as $0.3<(g-i)_0\leq 0.5$, and ``G stars'', defined as $0.5<(g-i)_0\leq 0.75$.
We adopt for these absolute magnitudes $M_g=3.5$ and $M_g=5.5$, respectively.
It is important to keep in mind that these assumptions apply only
to the putative ejectee population: the photometric properties
of non-ejectee stars, which obviously dominate the SDSS sample overall,
are completely irrelevant.

 From the observed dereddened magnitude $g_0$ of the star, we
can then infer its distance, $r=10^{(g_o-M_g)/5 - 2}\,$kpc,
and hence (from its position on the sky), infer the angle
$\phi$ between the Sun and the GC, as seen from the star.
Then, since the ejectee is assumed to be on a radial orbit, its full
3-D velocity is $v=|v_{r,G}\sec\phi|$, and hence its velocity
at the solar circle (assuming, as is always the case  for F stars
in SDSS, that $R<R_b$) is
\begin{equation}
v_{\odot\rm-circle} = \sqrt{(v_{r,G}\sec\phi)^2 + 2v_c^2\ln(R/R_0)}.
\label{eqn:sol_circ}
\end{equation}
It is immediately obvious that this equation places strong constraints
on the observational parameter space in which the search can be 
conducted.  For instance, if the star is fairly bright, so that 
$r\ll R_0$, then $\cos\phi>v_\cut/v_\esc$.  Since, for this geometry,
$\cos\phi\sim -\cos l\cos b$, this constraint eliminates the roughly 
$v_\cut/v_\esc= 73\%$ of the sky that is not sufficiently close to
the Galactocentric (or anticentric) directions.  At the opposite
extreme, a bound star at $R>R_0\exp[(v_\esc/v_\cut)^2/2]\sim 2.57\,R_0$
cannot be probed in any direction because its velocity (and hence
RV) will fall below the threshold.  At intermediate distances,
these two effects combine with different relative weights.

\subsection{{Candidate Bound Ejectees}
\label{sec:bound_candidates}}

Of the stars in the Knapp et al. (2009) sample, four F stars (and no G
stars) survive as candidate bound ejectees (see Table~\ref{tbl-1}).
It is a curious fact that all four are coming toward us, however there
is a 1/8 random probability that all four would be going in the same
direction, which is too high to warrant further analysis.  It is also
interesting that these stars are all measured to be metal poor.  These
candidates could either be members of a metal-poor population ejected from
the GC or simply halo stars. Obtaining proper motion measurements for
the candidates would discriminate between the two scenarios.

\subsection{{General Formulae for Bound-Ejectee Sensitivity}
\label{sec:bound_sensitivity}}

Initially, consider a spectroscopic survey that obtains RVs of
all stars in a small angular area $\Omega$ over a narrow magnitude
range $g_0\pm \Delta g_0/2$, and over a specified narrow range
of colors.  Consider a star in this sample that is ejected
from the GC, with current Galactocentric RV $v_r$, and
assume that it has an absolute magnitude (estimated from its color)
$M_g$.  The star then has distance from us $r= 10^{(g_0-M_g)/5 - 2}\,$kpc,
so the volume of such a survey is 
\begin{equation}
\Delta V = {\ln 10\over 5}\Delta g_0\Omega r^3.
\label{eqn:delta_vol}
\end{equation}
As mentioned above it has a 3-D Galactocentric velocity $v = v_r\sec\phi$,
where $\phi$ is the angle between us and the GC, as seen from the
star.  

As seen from the GC, the probed volume covers an
area $\Delta A$ and has a thickness $\Delta w$.  Of course,
$\Delta V = \Delta A\Delta w$. A fraction $\Delta A/4\pi R^2$
of all ejected stars will pass through this volume at some time
(assuming isotropic ejection),
where $R$ is the star's Galactocentric distance, and it will
spend a fraction $2\Delta w/vP$ of its time in this volume,
where $P$ is the orbital period.
Hence, observations of this volume will probe a fraction 
\begin{equation}
f_{\rm vol}={\Delta A\over 4\pi R^2}{2\Delta w\over vP}
= {\ln 10\over 2.5}{\Delta g_0\Omega r^3\over 4\pi R^2 vP}
\label{eqn:f_full}
\end{equation}
of all ejected stars. Now consider that we do not observe all stars in this volume,
but only one. Clearly, the fraction of ejected stars that
we probe falls by a factor $N_{\rm vol}$, where $N_{\rm vol}$ is 
the number of stars that satify our detection criteria.  That
is, $N_{\rm vol} = n\Omega\Delta g_0$, where $n$ is the number density
of stars per sterradian, per magnitude, and satisfying the color
criteria.  Hence, the fraction of all ejected stars probed by this
single observation of the $i$th star is
\begin{equation}
f_i= {\ln 10\over 2.5}{ r_i^3\over 4\pi R_i^2 v P_i n_i}.
\label{eqn:f_i}
\end{equation}

To determine the period, we apply our adopted Galactic model, first
finding that the apocenter of the orbit $R_m$ will be 
\begin{equation}
R_m = R_*,\quad (R_*\leq R_b);\qquad
R_m = {R_b\over 1 + \ln(R_b/R) - (v/v_c)^2/2} \quad (R_*>R_b)
\label{eqn:apocenter}
\end{equation}
where
\begin{equation}
R_* \equiv R\exp{(v/v_c)^2\over 2}.
\label{eqn:isothermal_apocenter}
\end{equation}
The velocity $v'$ at any point in the orbit $R'$ will be given by
\begin{equation}
{1\over 2}\biggl({v'\over v_c}\biggr)^2 = \ln{R_*\over R'},
\quad (R'\leq R_b);\qquad
{1\over 2}\biggl({v'\over v_c}\biggr)^2 = {R_b\over R'}- {R_b\over R_m},
\quad (R'> R_b).
\label{eqn:v_of_R}
\end{equation}
Integrating the equation $d t = d R'/v'(R')$ over one full period yields
\begin{equation}
P = \sqrt{2\pi}{R_*\over v_c},\qquad (R_*\leq R_b),
\label{eqn:period1}
\end{equation}
\begin{equation}
P = \sqrt{2\pi}{R_*\over v_c}{\rm erfc}\,[\sqrt{\ln(R_*/R_b)}]
+\sqrt{2}{R_m\over v_c}
\biggl[\sqrt{1-x} + {\arccos\sqrt{x}\over \sqrt{x}} \biggr]
,\qquad (R_*>R_b)
\label{eqn:period2}
\end{equation}
where $x\equiv R_b/R_m$.

The total sensitivity of the survey is then simply the sum $f_\tot
\equiv \sum_i f_i$ over all stars.  If there are $N_\tot$ bound stars
orbiting, then the expected number of detections will simply be
$N_{\rm exp} = f_\tot N_\tot$.  It will prove useful later on to
express the same result as
\begin{equation}
\Gamma_\thresh^{-1} =\tau_{\rm MW}\sum_i f_i,
\label{eqn:gamma_thresh}
\end{equation}
where $\tau_{\rm MW}=10\,$Gyr is the lifetime of the Milky Way.
In this formulation, $N_{\rm exp} = \Gamma/\Gamma_{\rm thresh}$,
where $\Gamma$ is the mean bound-ejection rate averaged over
the lifetime of the Galaxy.

To understand the basic properties of $f_i$, let us initially restrict
attention to the first case, $R_*\leq R_b$.  Then
\begin{equation}
f_i= {\ln 10\over 5}(2\pi)^{-3/2}
\biggr({r_i\over R_i}\biggr)^3{\exp(-z^2/2)\over z n_i},
\qquad z\equiv {v\over v_c}
\label{eqn:f_i_simple}
\end{equation}
If we consider some typical values for relatively faint turnoff stars in
SDSS ($g_0\sim 18.5$),
$r\sim 10\,$kpc, $R\sim 16\,$kpc,
$z=2$, and $n=100\,\rm deg^{-2}$, then $f_i\sim 1.5\times 10^{-9}$.  
Hence, by obtaining RVs for about $10^4$ stars, one could probe
a bound population of F stars ejected from the GC provided it
had at least $10^5$ members.  This corresponds to 
$\Gamma_\thresh= 10\,\rm Myr^{-1}$.

\subsection{{Bound-Ejectee Sensitivity of SDSS}
\label{sec:bound_sensitivity_sdss}}

\begin{figure}
\plotone{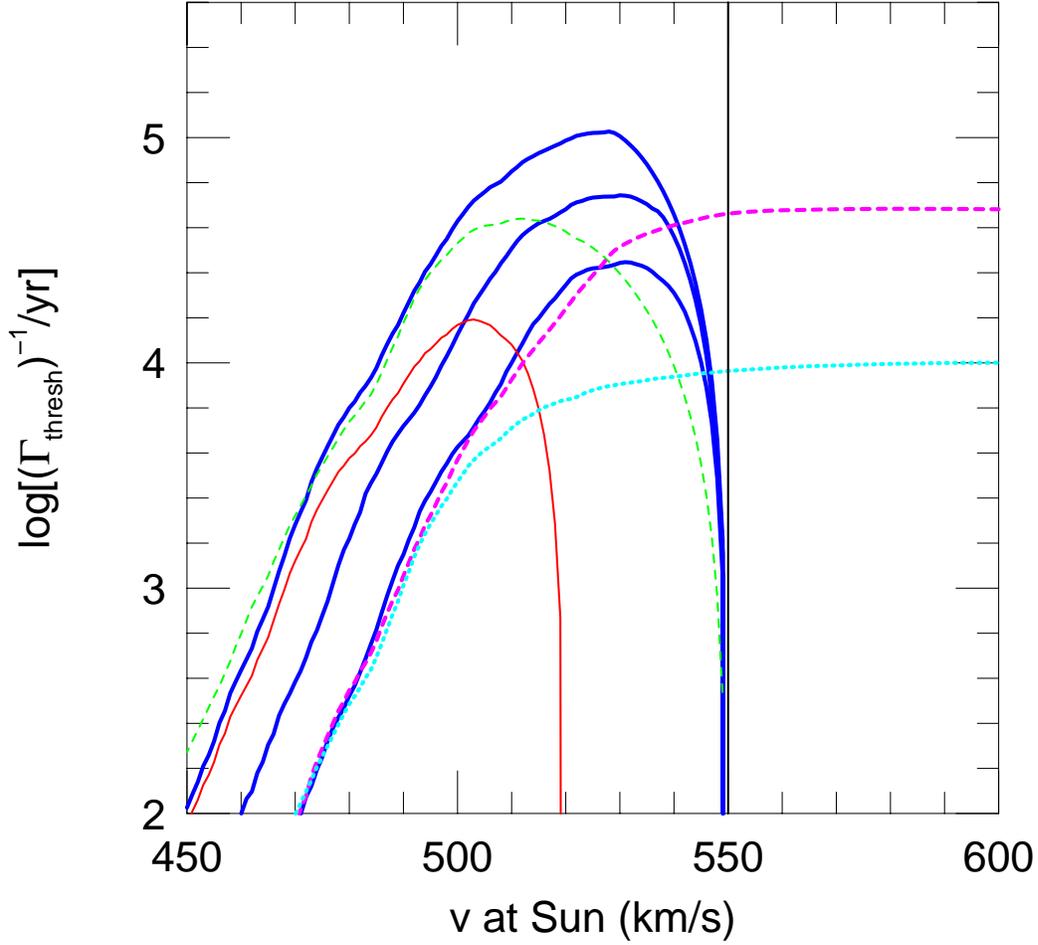}
\caption{\label{fig:sens} Sensitiviy of SDSS to old-star ejectees as a
  function of type and Galactic model.  Bold, blue curves show the
  sensitivity for bound F stars assuming $v_\esc=550\,\kms$ and for
  three different thresholds, $v_\cut = 400,$ 410, and 420 from bottom
  to top.  Solid red curve shows the case for bound F stars with
  $v_\cut=400\,\kms$, $v_\esc=520\,\kms$, while the dashed (green)
  curve is for bound G stars with $v_\cut=400\,\kms$,
  $v_\esc=550\,\kms$.  The sensitivity to unbound F stars (bold
  dashed, magenta) and unbound G stars (dotted, cyan) for the case
  $v_\esc = 550\,\kms$, $v_\cut = 400\,\kms$ are also shown.  }\end{figure}

To determine the sensitivity of the SDSS spectroscopic survey, we
first evaluate $n_i$ for our two classes of stars as a function of
magnitude and position on the sky using the SDSS photometric catalog.
This was done using 20 individual patches of sky, each with area 1
square degree, and interpolating these values to obtain $n_i$ for
other parts of the sky.  We then sum equation (\ref{eqn:f_i}) over all
stars in the SDSS spectroscopic catalog that meet our color criteria.
Figure \ref{fig:sens} shows the result for several cases.  The three
bold curves (blue) are for F stars under the assumption of
$v_\esc=550\,\kms$ and for three different thresholds, $v_\cut = 400,$
410, and 420.  The solid (red) curve is for F stars with
$v_\cut=400\,\kms$, $v_\esc=520\,\kms$, while the dashed (green) curve
is for G stars with $v_\cut=400\,\kms$, $v_\esc=550\,\kms$.  The two
other curves will be explained later.

Several features are apparent from this figure.  First, sensitivity is
peaked quite close to the escape velocity.  This may seem surprising
in view of equation (\ref{eqn:f_i_simple}), which exponentially
declines with the square of the velocity for $R_*<R_b$, and then even
more steeply.  However, at low $v$, distant (i.e., faint) stars do not
contribute because their declining speeds at large $R$ bring them
below the first selection cut in equation (\ref{eqn:sel_crit}).  Most
distant stars gain by $r^3$ in equation (\ref{eqn:f_i_simple}), at
least for $r<R_0$.  Second, sensitivity falls quite steeply with
increasing $v_\cut$, approximately a factor 2 for each $10\,\kms$.
This is basically a consequence of the same volume effect just
analyzed.  The same effect also accounts for the dramatic falloff in
sensitivity if the escape velocity proves lower than our default
$v_\esc=550\,\kms$.  Finally, senstivity to bound G-star ejectees is
about 1/3 that of F stars.

\begin{figure}
\plotone{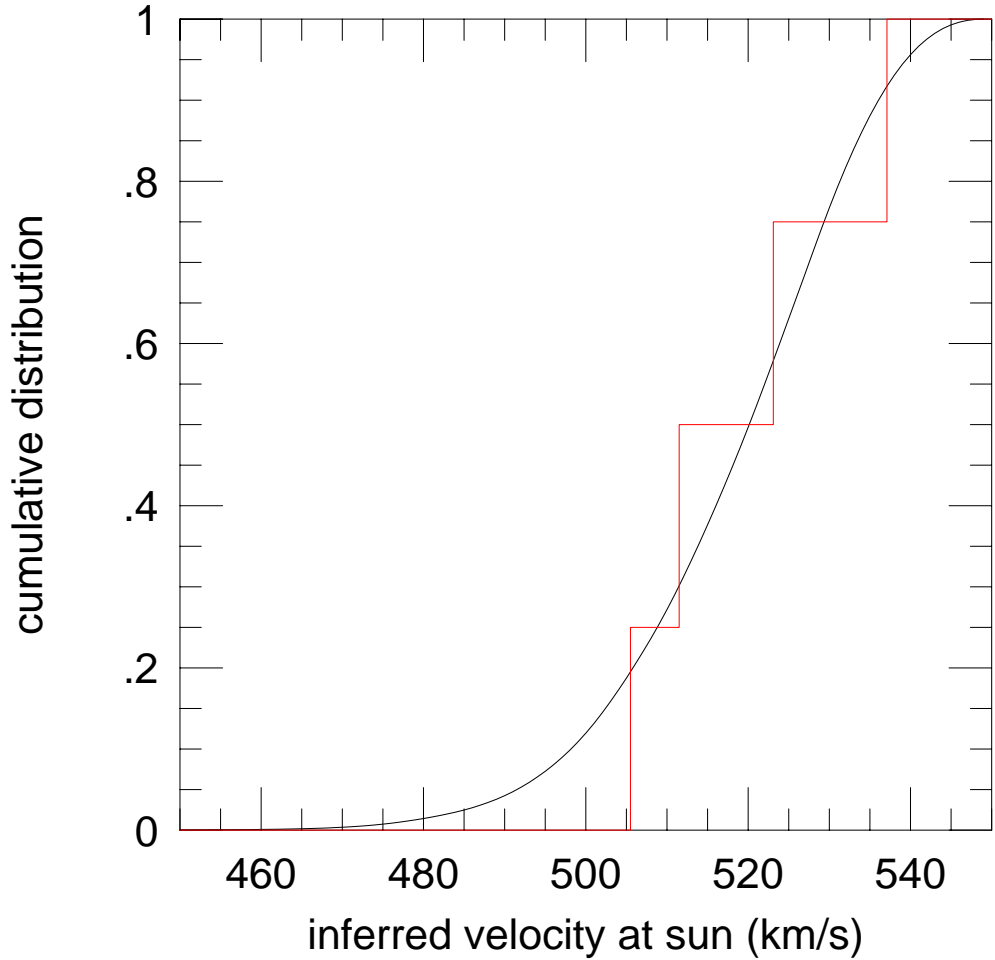}
\caption{Cumulative distribution of velocities of candidate bound F-star HVS population (red histogram).  A simple model in which the number of bound F-star ejectees is a uniform function of $v_{\odot \rm-circle}$ is shown in black for comparison. \label{fig:cum}
}\end{figure}

Let us initially adopt the extremely simple model
that the number of bound F-star ejectees is uniform as a function
of $v_{\odot -\rm circle}$.  Under this assumption, we can make
one test of the plausibility of the candidate population by
ploting their cumulative distribution, which is shown in
Figure \ref{fig:cum}.  The match is extremely good.

Let us then tentatively assume that the candidates are really 
bound ejectees.  Integrating under the entire (upper-blue) curve in Figure \ref{fig:sens}
yields $\int d v\Gamma_\thresh^{-1} = 4.0\,\rm Myr$-$\kms$.  However, from
Figure \ref{fig:sens} and especially Figure \ref{fig:cum}, it is clear
that about 75\% of this sensitivity comes from the roughtly
$35\,\kms$ interval in which the four candidates are detected.
Therefore, if these candidates are accepted as real, we can
infer an ejection rate
\begin{equation}
\Gamma \sim {4\over 
\langle\Gamma_{\rm thresh}^{-1}\rangle} = 
4{35\,{\rm \kms}\over 0.75 \times 4\,{\rm Myr}^{-1}{\rm-}\kms}\sim 45\,\rm Myr^{-1} 
\qquad (505\,\kms < v_{\odot\rm-circle}<540\,\kms)
\label{eqn:gamma_eval}
\end{equation}
Of course, if some or all of the candidates are later vetted and found
wanting, the smaller number of detections would then yield an
analagous upper limit.

Since the G-star sensitivity is smaller and we find no candidates, the
G stars are consistent with a similar rate, but with weaker
statistics.

\subsection{{Unbound Ejectees}
\label{sec:unbound_ejectees}}

We select candidate unbound ejected stars with exactly the criterion as
bound stars (eq.~[\ref{eqn:sel_crit}]), except reversing the
sign on the second condition, $v_{\odot\rm -circle}\geq v_\esc$, and
of course demanding that they be exiting rather than entering the Galaxy.
Then, following a very similar derivation to the one given for 
equations (\ref{eqn:f_i}) and (\ref{eqn:gamma_thresh}), 
we immediately derive their analogs for the unbound case,
\begin{equation}
\Gamma_\thresh^{-1} = \sum_i t_i
\qquad
t_i= {\ln 10\over 5}{ r_i^3\over 4\pi R_i^2 v n_i}.
\label{eqn:t_i}
\end{equation}
This quantity is shown in Figure \ref{fig:sens}
as a function of $v_{\odot\rm -circle}$ for F stars
(bold dashed, magenta) and G stars (dotted, cyan) for
the case $v_\esc = 550\,\kms$, $v_\cut = 400\,\kms$.
Nothing prevents us from evaluating equation (\ref{eqn:t_i})
even below $v_\esc$, and we display the extensions of these
curves in Figure \ref{fig:sens}.  Their physical meaning is that
it is possible to detect bound ejectees during their first orbit,
and for orbital times of order or longer than $\tau_{\rm MW}$,
this is actually a more accurate representation of the detection
rate than averaging over a full orbits. Thus,
the bound rate is really the maxium of the bound curve and
extension of the unbound curve.

There is one F star candidate that survives our selection criteria and
one G star.  The proper motions for both of these stars are
inconsistent with a Galactocentric origin as we now show.  Were it
ejected from the GC, based on its current position, a star should
exhibit a proper motion in the direction
\begin{equation}
{\rm tan}(\theta) = {\rm tan}(l){\rm sin}(b)
\end{equation}
where $\theta$ is measured from Galactic north through east.  The F
star candidate has a measured proper motion from SDSS \citep{pier03,munn04}
of $34\,{\rm mas\, yr^{-1}}$ at $55^\circ$ (with small errors) ---
significantly in conflict with the predicted value $\theta=330^\circ$.

Unlike the F star, the G star candidate cannot be discarded based on
the its proper-motion direction alone. The measured proper motion of
$6\,{\rm mas\, yr^{-1}}$ is sufficiently small that the measured
direction of $183^{\circ}$ is poorly determined given the errors.  We
therefore compare the predicted and observed {\it vector} proper
motions, which (unlike the case of the F star) require a distance
estimate.  From its $g_0$ magnitude we estimate a distance of
$7.8\kpc$, and it is therefore expected to be moving at 14$\,{\rm
  mas\, yr^{-1}}$ at $244^{\circ}$.  The magnitude of the vector
difference of the expected and observed proper motions is $12\,{\rm
  mas\, yr^{-1}}$, which is to be compared to the typical errors of
$\sim 4\,{\rm mas\, yr^{-1}}$ \citep{munn04}, making it unlikely that
this star is an unbound ejectee.

Since $\Gamma_\thresh$ is essentially constant in the unbound regime
at (0.05 Myr)$^{-1}$ for F stars and (0.01 Myr)$^{-1}$ for G stars,
this lack of detections places an upper limit on the ejection of
unbound F and G stars of
\begin{equation}
\Gamma^F_{\rm unbound} < 60\,\rm Myr^{-1},\qquad
\Gamma^G_{\rm unbound} < 300\,\rm Myr^{-1},
\label{eqn:unbound_f_g}
\end{equation}
at 95\% confidence.

Equation (\ref{eqn:unbound_f_g}) is broadly consistent with equation
(\ref{eqn:gamma_eval}).

\section{{Comparison with B stars}
\label{sec:comp_b_stars}}

The Brown et al. (2007) survey in essence covered an entire volume
of Galaxy rather than a scattering of targets within the Galaxy,
like SDSS.  Thus, to calculate $\Gamma_\thresh$ one should
perform a volume integral.  An argument similar to the one
given in \S~\ref{sec:bound_sensitivity} leads to the equation
\begin{equation}
\Gamma_\thresh^{-1} = \int_{\rm area-covered} {d\sin b d l\over 4\pi}
\int_{r_{\rm min}}^{r_{\rm max}}
{d r\,r^2\over [R(r,l,b)]^2[v^2 - 2v_c^2\ln(R/R_0)]^{1/2}}
\label{gamma_brown}
\end{equation}
where we have for convenience assumed the form of the Galactic potential
within $R<R_b$.  We esitmate this expression for the
Brown et al. (2007) survey by noting that it covered an area 
$\Delta\Omega=5000\,\rm deg^{2}$ and that most of the sensitivity
was in regions where $R\sim r$, by adopting an average value
for the velocity in the denominator of $\langle v\rangle = 600\,\kms$,
and approximating $\Delta r = r_{\rm max} - r_{\rm min}= 40\,$kpc.
We then obtain 
$\Gamma_\thresh^{-1} = (\Delta\Omega/4\pi)(\Delta r/\langle v\rangle) =
8\,\rm Myr$.  Hence the four detections imply 
$\Gamma^B_{\rm unbound} = 4\Gamma_\thresh = 0.5\,\rm Myr^{-1}$.

Thus, the upper limit given by equation (\ref{eqn:unbound_f_g})
shows that F-star ejectees are no more than 100 times more common
than B-star ejectees.  Since F stars themselves are about 100 times
more common than B stars in the Galaxy as a whole, this result
is already of some interest.  Nevertheless, it would certainly
be better to achieve at least a few times better sensitivity.

\section{Discussion \label{sec:disc}}

We have presented the first limit on old-population ejectees from the
Galactic Center based on data obtained with the SDSS.  We have shown
that the ejection of F/G stars is no more than 100 times more common
than O/B star ejectees.  We note that this limit is obtained simply
from analyzing the stellar spectra obtained by SDSS as part of its
main projects, {\it not} via a survey designed to preferentially
select these objects.  Already, this limit suggests that the mass
function at the Galactic center is not weighted toward low mass star
formation --- a fact impossible to determine by direct imaging of the
GC alone.  Because these candidates are all found to be metal poor
(see Table~\ref{tbl-1}), they may be part of the metal-poor halo in
which case, this limit is even stronger.  Proper motion measurements
would determine whether they are indeed ejectees from the GC and
thereby make this limit more stringent.  Ongoing surveys, such as the
second epoch of the Sloan Extension for Galactic Understanding and
Exploration (SEGUE-II), will determine conclusively if star-formation
at the GC is indeed weighted heavily toward young, massive stars.

\acknowledgements

We thank Warren Brown for helpful comments on this manuscript.  JAK
was supported by Hubble Fellowship HF-01197 and by a
Carnegie-Princeton Fellowship during this work.  AG was supported by
NSF grant AST-0757888. JAK and AG acknowledge the support of the Kavli
Institute for Theoretical Physics in Santa Barbara during the
completion of this work.  T.C.B. acknowledges partial funding of this
work from grants PHY 02-16783 and PHY 08-22648: Phyiscs Frontiers
Center/Joint Institute for Nuclear Astrophics (JINA), awarded by the
U.S. National Science Foundation.  This research has made use of
NASA's Astrophysics Data System, of the SIMBAD data base operated at
CDS, Strasbourg, France, and of the plotting and analysis package SM,
written by Robert Lupton and Patricia Monger.  Funding for the
creation and distribution of the SDSS-I and SDSS-II Archives has been
provided by the Alfred P. Sloan Foundation, the Participating
Institutions, the National Aeronautics and Space Administration, the
National Science Foundation, the U.S. Department of Energy, the
Japanese Monbukagakusho, the Max Planck Society, and the Higher
Education Funding Council of the United Kingdom. The SDSS Web site is
http://www.sdss.org/. The SDSS is managed by the Astrophysical
Research Consortium (ARC) for the Participating Institutions. The
Participating Institutions are The American Museum of Natural History,
Astrophysical Institute Potsdam, University of Basel, Cambridge
University, Case Western Reserve University, The University of
Chicago, The Chinese Academy of Sciences (LAMOST), Drexel University,
Fermilab, the Institute for Advanced Study, the Japan Participation
Group, The Johns Hopkins University, the Joint Institute for Nuclear
Astrophysics, The Kavli Institute for Particle Physics and Cosmology,
the Korean Scientist Group, Los Alamos National Laboratory, the
Max-Planck-Institute for Astronomy (MPIA), the Max-Planck-Institute
for Astrophysics (MPA), New Mexico State University, The Ohio State
University, University of Pittsburgh, University of Portsmouth,
Princeton University, the United States Naval Observatory, and the
University of Washington.

\clearpage
\begin{deluxetable}{llccccccll}
\tabletypesize{\scriptsize}
\tablenum{1}
\tablecolumns{9}
\tablecaption{Candidate Ejectees Satisfying Velocity Criteria \label{tbl-1}}
\tablehead{
\colhead{Name}&
\colhead{$V_G$~(km/sec)}&
\colhead{Type}&
\colhead{$g_0$}&
\colhead{$(g-i)_0$}&
\colhead{${\rm T_{eff}~(K)}$}&
\colhead{log(g)}&
\colhead{[Fe/H]}&
\colhead{Plate/Fiber/MJD}&\\
}
\startdata

SDSSJ061118.63+642618.5&  -412.3$\pm$5.5&  F& 18.39& 0.45 & 6143 & 4.08 & -1.27 & 2299/488/53711 & \\
SDSSJ074557.31+181246.7&  -409.0$\pm$5.9&  F& 18.96& 0.34 & 6235 & 3.51 & -1.73 & 2074/113/53437 & \\
SDSSJ224052.56+011332.1&  -407.1$\pm$16.3& F& 19.30& 0.49 & 6168 & 4.02 & -1.37 & 1101/561/52621 & \\
SDSSJ211321.02+103456.7&  -404.4$\pm$17.2& F& 20.07& 0.35 & 6069 & 4.10 & -1.91 & 1891/423/53238 & \\
SDSSJ111107.85+585357.2&  +425.4$\pm$3.1&  F& 16.16& 0.36 & 6514 & 3.58 & -1.77 & 950/554/52378  &\\
SDSSJ224740.09-004451.6&  +401.8$\pm$13.1& G& 19.97& 0.72 & 5518 & 4.17 & -1.22 & 1901/7/53261   &\\ 

\enddata

\end{deluxetable}
\clearpage

\end{document}